\documentclass[aps,
groupedaddress,
reprint,
superscriptaddress,
nofootinbib,
nobibnotes,
amsmath,amssymb,
 prd,
 10pt,
 ]{revtex4-2}
\usepackage{graphicx}
\usepackage{bm}
\usepackage{hyperref}
\usepackage{booktabs}
\usepackage{lineno}

\makeatletter
\renewcommand{\@fnsymbol}[1]{\ifcase#1\or a\or *\or c\or d\or e\fi}
\makeatother

\begin{document}

\title{DREAMuS: Dark matter REsearch with Advanced Muon Source} 

\author{Xiang Chen}
\affiliation{State Key Laboratory of Dark Matter Physics, School of Physics and Astronomy, Shanghai Jiao Tong University, Shanghai, China}
\affiliation{Key Laboratory for Particle Astrophysics and Cosmology (Ministry of Education), Shanghai Jiao Tong University, Shanghai, China}
\affiliation{Shanghai Key Laboratory for Particle Physics and Cosmology, Shanghai Jiao Tong University, Shanghai, China}

\author{Zejia Lu}
\affiliation{State Key Laboratory of Dark Matter Physics, School of Physics and Astronomy, Shanghai Jiao Tong University, Shanghai, China}
\affiliation{Key Laboratory for Particle Astrophysics and Cosmology (Ministry of Education), Shanghai Jiao Tong University, Shanghai, China}
\affiliation{Shanghai Key Laboratory for Particle Physics and Cosmology, Shanghai Jiao Tong University, Shanghai, China}

\author{Liangwen Chen}
\affiliation{Advanced Energy Science and Technology Guangdong Laboratory, Huizhou 516000, China}
\affiliation{Institute of Modern Physics, CAS, Lanzhou 730000, China}
\affiliation{School of Nuclear Science and Technology, University of Chinese Academy of Sciences, Beijing 100049, China}

\author{Jun Gao}
\affiliation{State Key Laboratory of Dark Matter Physics, School of Physics and Astronomy, Shanghai Jiao Tong University, Shanghai, China}
\affiliation{Key Laboratory for Particle Astrophysics and Cosmology (Ministry of Education), Shanghai Jiao Tong University, Shanghai, China}
\affiliation{Shanghai Key Laboratory for Particle Physics and Cosmology, Shanghai Jiao Tong University, Shanghai, China}

\author{Shao-Feng Ge}
\affiliation{State Key Laboratory of Dark Matter Physics, Tsung-Dao Lee Institute and School of Physics and Astronomy, Shanghai Jiao Tong University, Shanghai, China}
\affiliation{Key Laboratory for Particle Astrophysics and Cosmology (Ministry of Education), Shanghai Jiao Tong University, Shanghai, China}
\affiliation{Shanghai Key Laboratory for Particle Physics and Cosmology, Shanghai Jiao Tong University, Shanghai, China}

\author{Zhanxu Hao}
\affiliation{State Key Laboratory of Dark Matter Physics, School of Physics and Astronomy, Shanghai Jiao Tong University, Shanghai, China}
\affiliation{Key Laboratory for Particle Astrophysics and Cosmology (Ministry of Education), Shanghai Jiao Tong University, Shanghai, China}
\affiliation{Shanghai Key Laboratory for Particle Physics and Cosmology, Shanghai Jiao Tong University, Shanghai, China}

\author{Yang Hu}
\affiliation{State Key Laboratory of Dark Matter Physics, School of Physics and Astronomy, Shanghai Jiao Tong University, Shanghai, China}
\affiliation{Key Laboratory for Particle Astrophysics and Cosmology (Ministry of Education), Shanghai Jiao Tong University, Shanghai, China}
\affiliation{Shanghai Key Laboratory for Particle Physics and Cosmology, Shanghai Jiao Tong University, Shanghai, China}

\author{Bingzhi Li}
\affiliation{Scientific Model Research Group, Zhejiang Lab, Hangzhou, China}

\author{Cen Mo}
\affiliation{State Key Laboratory of Dark Matter Physics, School of Physics and Astronomy, Shanghai Jiao Tong University, Shanghai, China}
\affiliation{Key Laboratory for Particle Astrophysics and Cosmology (Ministry of Education), Shanghai Jiao Tong University, Shanghai, China}
\affiliation{Shanghai Key Laboratory for Particle Physics and Cosmology, Shanghai Jiao Tong University, Shanghai, China}

\author{Zhiyu Sun}
\affiliation{Advanced Energy Science and Technology Guangdong Laboratory, Huizhou 516000, China}
\affiliation{Institute of Modern Physics, CAS, Lanzhou 730000, China}
\affiliation{School of Nuclear Science and Technology, University of Chinese Academy of Sciences, Beijing 100049, China}

\author{Huayang Wang}
\affiliation{State Key Laboratory of Dark Matter Physics, School of Physics and Astronomy, Shanghai Jiao Tong University, Shanghai, China}
\affiliation{Key Laboratory for Particle Astrophysics and Cosmology (Ministry of Education), Shanghai Jiao Tong University, Shanghai, China}
\affiliation{Shanghai Key Laboratory for Particle Physics and Cosmology, Shanghai Jiao Tong University, Shanghai, China}

\author{Chonghao Wu}
\affiliation{State Key Laboratory of Dark Matter Physics, School of Physics and Astronomy, Shanghai Jiao Tong University, Shanghai, China}
\affiliation{Key Laboratory for Particle Astrophysics and Cosmology (Ministry of Education), Shanghai Jiao Tong University, Shanghai, China}
\affiliation{Shanghai Key Laboratory for Particle Physics and Cosmology, Shanghai Jiao Tong University, Shanghai, China}

\author{Yu Xu}
\affiliation{Advanced Energy Science and Technology Guangdong Laboratory, Huizhou 516000, China}
\affiliation{Institute of Modern Physics, CAS, Lanzhou 730000, China}

\author{Xueheng Zhang}
\affiliation{Advanced Energy Science and Technology Guangdong Laboratory, Huizhou 516000, China}
\affiliation{Institute of Modern Physics, CAS, Lanzhou 730000, China}
\affiliation{School of Nuclear Science and Technology, University of Chinese Academy of Sciences, Beijing 100049, China}

\author{Yulei Zhang}
\altaffiliation{Previously at School of Physics and Astronomy, Shanghai Jiao Tong University, Shanghai 200240, China}
\affiliation{Department of Physics, University of Washington, Seattle, Washington, USA}

\author{Liang Li}
\email{liangliphy@sjtu.edu.cn}
\affiliation{State Key Laboratory of Dark Matter Physics, School of Physics and Astronomy, Shanghai Jiao Tong University, Shanghai, China}
\affiliation{Key Laboratory for Particle Astrophysics and Cosmology (Ministry of Education), Shanghai Jiao Tong University, Shanghai, China}
\affiliation{Shanghai Key Laboratory for Particle Physics and Cosmology, Shanghai Jiao Tong University, Shanghai, China}

\date{\today}

\begin{abstract}
We propose DREAMuS, a fixed-target experiment at the High Intensity Heavy-Ion Accelerator Facility (HIAF), to search for muon-philic dark matter mediated by light flavor-violating bosons. DREAMuS is designed to probe the parameter space of a muon-philic dark matter (DM) mediated by a light flavor-violating boson, specifically
a vector \(Z'\) (or a scalar \(\phi\)) which is produced in muon--nucleus interactions and decays into dark matter particles with a distinctive detector signature. Precision tracking and time-of-flight measurements are used to suppress the Standard Model backgrounds. 
We find that DREAMuS can achieve competitive sensitivity in the GeV-scale muon-philic dark matter parameter space, reaching sensitivity to couplings at the $10^{-4}$, especially in the few-hundred-MeV region.
In addition to a \(\mu^-\) run, we highlight the potential of a complementary \(\mu^+\) beam option, further improving sensitivity to dark matter below 200 $\mathrm{MeV}$ by an order of magnitude.
\end{abstract}
\maketitle

\section{Introduction}

\label{sec:model}
Dark matter (DM) provides strong evidence for physics beyond the Standard Model (SM)~\cite{PhysRevD.79.015014}. Motivated by dark matter production via the thermal freeze-out mechanism, 
scenarios with light dark matter often feature a light mediator that connects the dark sector to the SM particles~\cite{Billard:2021uyg,Izaguirre:2014bca,Agrawal_2014}.
Such mediators are generically required to lie in the sub-GeV mass range and to be weakly coupled in order to remain consistent with existing experimental constraints~\cite{Bickendorf:2022buy}.

Light mediators coupled to leptons are well motivated. They have been invoked as a possible explanation of the muon anomalous magnetic moment discrepancy, \(a_\mu\equiv(g-2)_\mu/2\)~\cite{Leveille:1978fr,Pospelov:2008zw,Aoyama:2020ynm,Borsanyi:2020mff,Muong-2:2021vma,Muong-2:2023cdq,Muong-2:2025xyk}. We note that the updated 2025 Muon g-2 Theory Initiative white paper substantially reduces this tension~\cite{Aliberti:2025beg}, however unresolved tensions remain among data-driven dispersive evaluations, and a recent lattice-QCD calculation of the next-to-leading-order hadronic vacuum polarization finds a \(4.8\sigma\) discrepancy with data-driven evaluations excluding CMD-3~\cite{Beltran:2026ofp}.

One simple example is provided by gauge extensions based on the difference of muon and tau lepton numbers, \(L_\mu-L_\tau\)~\cite{He:1991qd, NA64:2024nwj, Altmannshofer:2016jzy}, which introduces a light vector boson \(Z'\) coupling to second and third-generation leptons. In this model, the mediator is produced via flavor-conserving radiation off the incoming muon hitting on target (N),
\(\mu + N \to \mu + N + Z'\),
which has been extensively studied by searches at fixed-target experiments~\cite{NA64:2024klw,Kahn:2018cqs, wang2025searchlightdarksectors}.

Beyond purely flavor-conserving interactions, light leptonic mediators may also induce charged-lepton flavor violation (LFV), enabling transitions such as \(\mu \to e\). 
While low-energy muon decay measurements provide constraints on flavor-violating couplings, the region where the mediator mass exceeds the muon mass remains unexplored.
Recent studies have shown that leptonic scalar portal models, in which a light scalar mediator carries both SM and dark-sector interactions, can simultaneously contribute to \((g-2)_\mu\) and connect to dark matter through predominantly invisible decays~\cite{Gninenko:2022ttd}. 

At the effective level, these scenarios are conveniently described by a neutral leptonic mediator \(X\), which may be either a vector boson $Z'_\mu$ or a scalar \(\phi\). The interaction Lagrangian can be written as:
\begin{equation}
\mathcal{L}\;\supset\;
g_\mu\,X\,\bar{f}\,\Gamma\,f
\;+\;
g_{e\mu}\,X\,\bar{e}\,\Gamma\,\mu
\;+\;\mathrm{h.c.},
\end{equation}
where \(\Gamma = \gamma^\mu\) for a vector mediator and \(\Gamma = 1\) for a scalar mediator.
The coupling \(g_\mu\) controls interactions with muons, while \(g_{e\mu}\) induces LFV processes. 

The model may further extend to include coupling with dark matter particle \(\chi\)\cite{Billard:2021uyg},
\begin{equation}
\mathcal{L} \supset g_{\chi}\, X\,\bar \chi \Gamma  \chi,
\end{equation}
allowing invisible decays $X\rightarrow\chi \bar{\chi}$. $g_{\chi}$ is the coupling of the $X$ to dark matter particles. 
Existing constraints for $X$ arise from LFV meson decays, rare muon decay processes, and collider searches ~\cite{CMS:2025wqy, ParticleDataGroup:2024cfk}, but a significant region of parameter space remains unexplored in the sub-GeV region.

\begin{figure}[htbp] 
\centering
\includegraphics[width=0.28\linewidth]{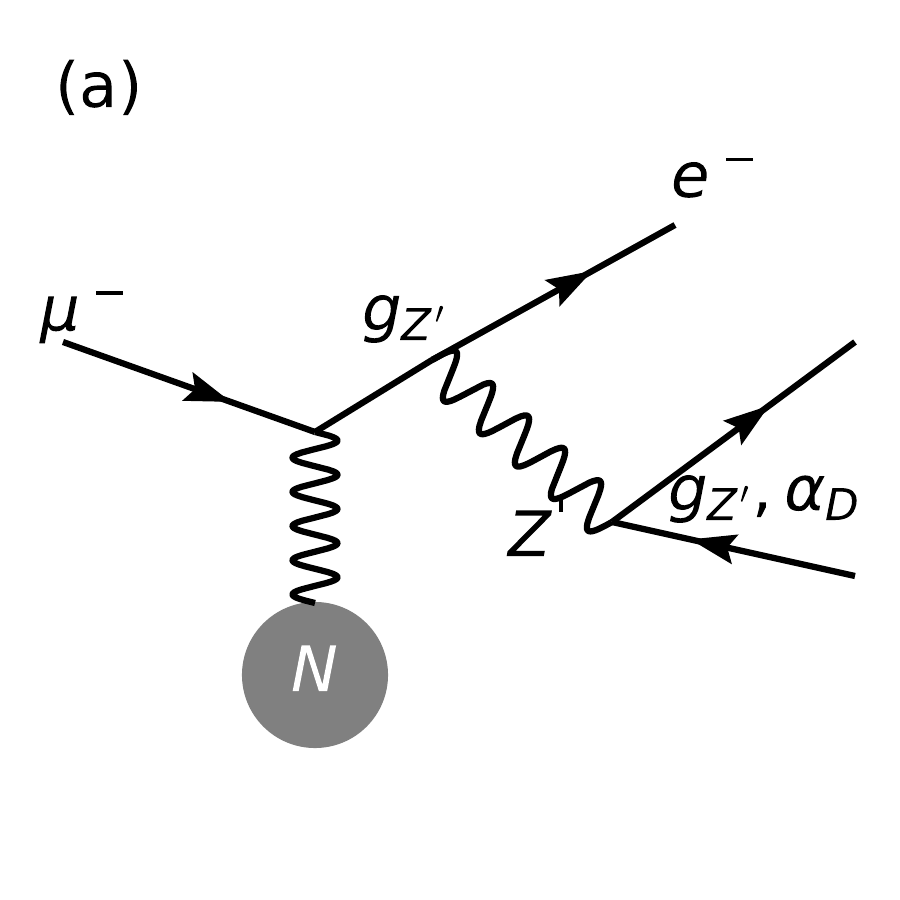} 
\includegraphics[width=0.28\linewidth]{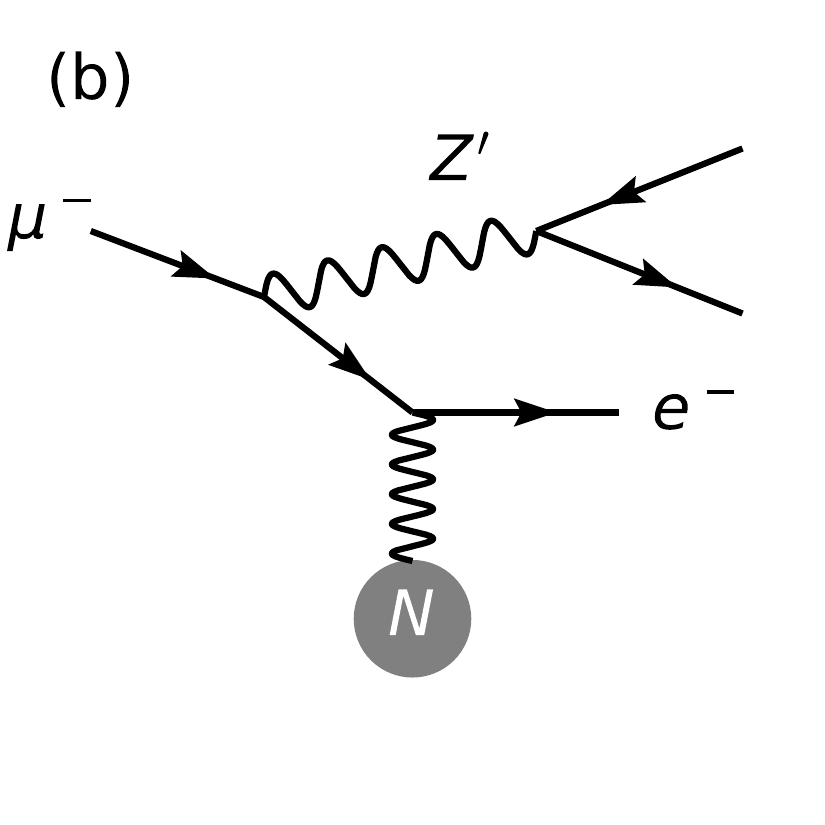} 
\includegraphics[width=0.28\linewidth]{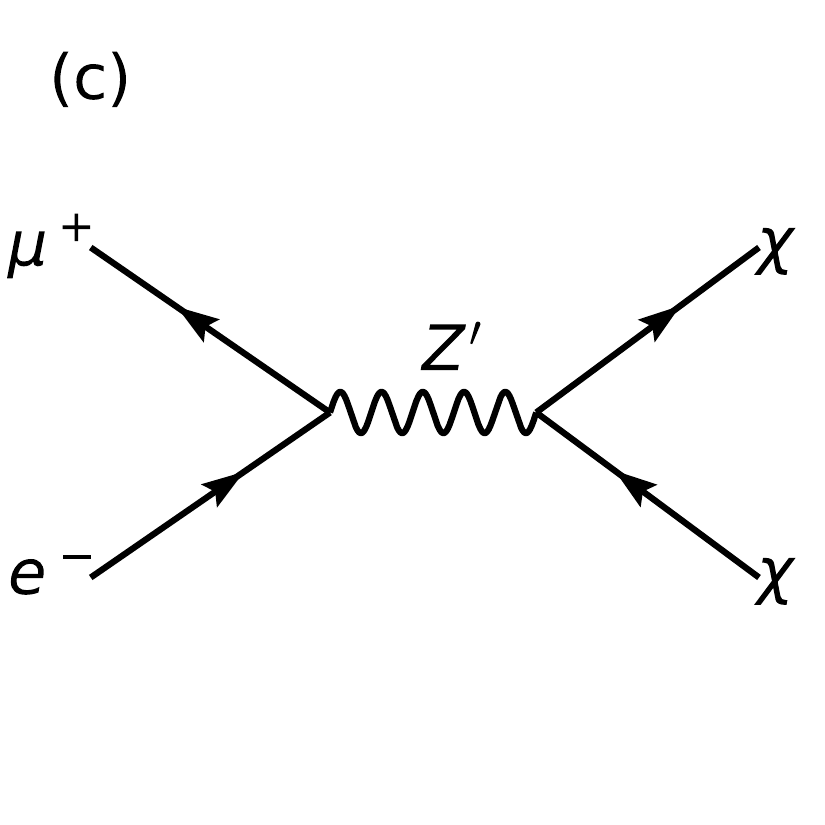} 
\includegraphics[width=0.28\linewidth]
{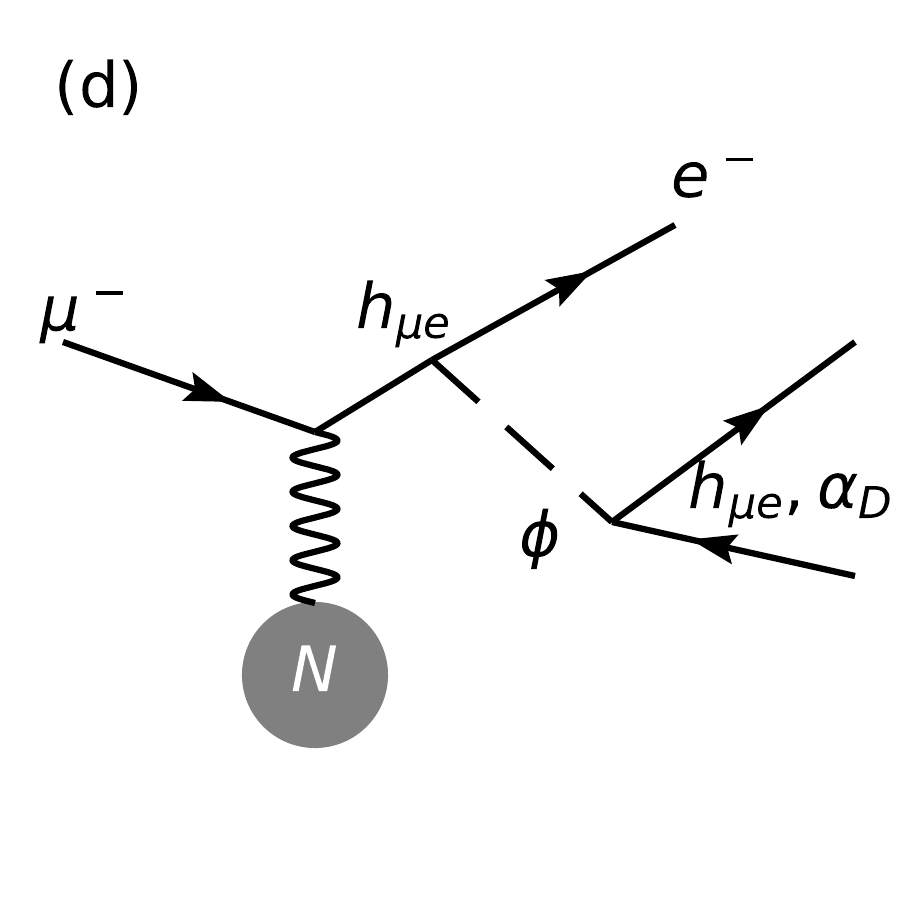} 
\includegraphics[width=0.28\linewidth]{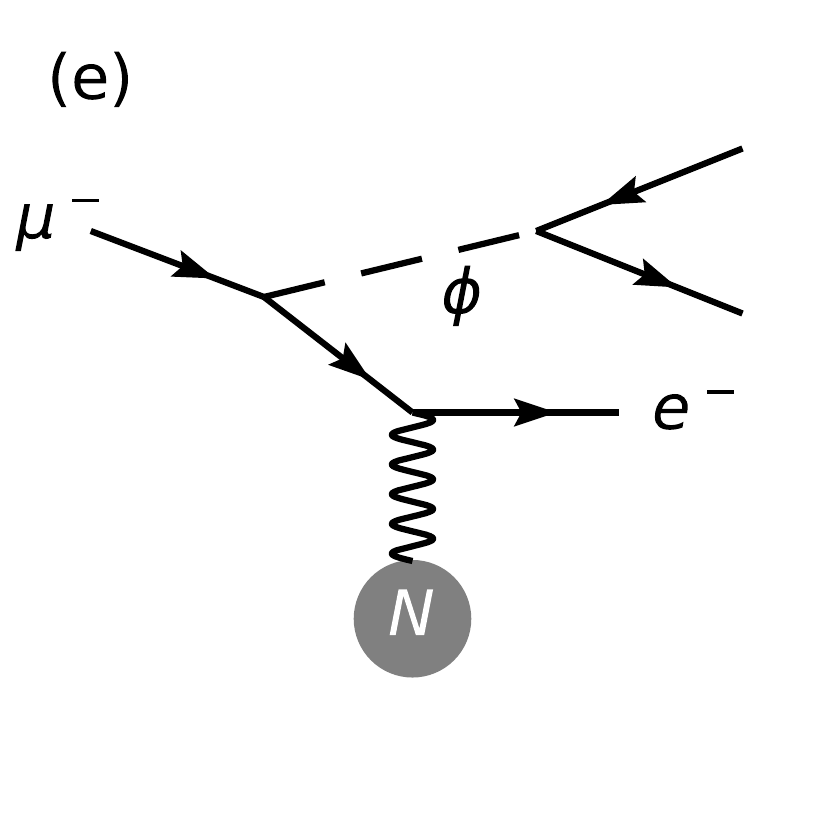}
\includegraphics[width=0.28\linewidth]{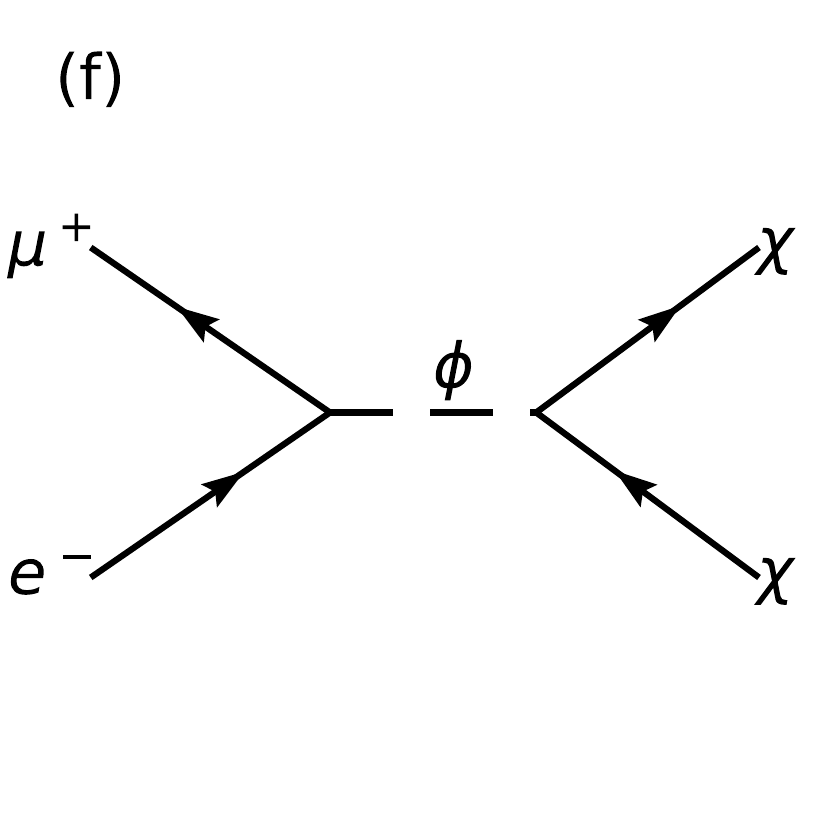}
\caption{Flavor-violating processes in a muon fixed-target experiment mediated by a vector $Z'$ (top row, panels a--c) or a scalar $\phi$ (bottom row, panels d--f).(c) and (f) represent the annihilation channel.} 
\label{fig:figure_feynman}
\end{figure}

High-intensity muon fixed-target experiments provide a sensitive probe of $Z'$ and $\phi$. In muon–nucleus scattering, the mediator can be radiated off the incoming muon in a bremsstrahlung-like process, which is referred as the radiation channel. The Feynman diagram is shown in Fig.~\ref{fig:figure_feynman}.
\begin{equation}
\mu^- + N \to e^- + N + X,\qquad X \to \chi\bar{\chi},
\end{equation}
leading to a final state with a single recoil electron and missing energy. The signal signature is clean, 
with a recoiled electron with sizable transverse momentum and no additional charged particles from the interaction vertex.

For positive muon beams, an additional production mode, the annihilation channel, arises from annihilation with atomic electrons in the target besides the radiation channel mentioned above,
\begin{equation}
\mu^+ + e^- \to X,\qquad X \to \chi\bar{\chi}.
\end{equation}
This final state is fully invisible. Compared to the radiation channel, the annihilation channel can provide enhanced sensitivity in the low-mass region through a resonance. 

In this work, we propose DREAMuS, a muon fixed-target experiment at the High Intensity Heavy-Ion Accelerator Facility (HIAF). The high-intensity muon beam available at HIAF enables a sensitive search for light leptonic mediators with flavor-violating couplings. We focus on the radiation channel with a single recoil electron and the annihilation channel. We show that with precise tracking and time-of-flight (TOF) measurements, SM backgrounds can be strongly suppressed, allowing DREAMuS to probe previously unexplored parameter space for muon-philic dark sectors. 

The paper is organized as follows. Section~2 describes the experimental setup and detector concept. Section~3 presents signal simulation and background estimation. Section~4 outlines the event selection strategy. Section 5 reports the projected sensitivity.

\section{Experimental setup}
\label{sec:setup}


The High Intensity Heavy-Ion Accelerator Facility (HIAF)~\cite{cpl_42_11_110102,Xu:2025spd} is a next-generation accelerator complex under construction in Huizhou, China. It is designed to deliver high-intensity ion beams at the GeV scale. By combining a superconducting ion linac, a booster synchrotron, and advanced fragment separation systems, HIAF provides a suitable platform for nuclear physics, hadron physics, and muon experiments.

A key component of HIAF is the high-energy fragment separation system (HIRIBL)~\cite{wang2025searchlightdarksectors} which enables the production and transport of high-momentum secondary particles. Through magnetic rigidity selection and optimized beam optics, HIRIBL can deliver muon beams with momenta up to approximately 7.5\,GeV. The beam momentum, intensity, and transverse spot size can be tuned to match different experimental requirements, allowing both high-rate and high-resolution measurements.

For a negative muon beam at around 3\,GeV, the HIRIBL can deliver intensities as high as $3.5\times10^{6}\,\mu/\mathrm{s}$ in continuous mode and $8.2\times10^{6}\,\mu/\mathrm{s}$ in pulsed mode~\cite{Xu:2025spd}. The beam can be operated in continuous mode by stretching each bunch from an initial width of about 400\,ns to approximately 3\,s. Compared with the pulsed mode, the beam intensity in continuous mode is reduced by about one order of magnitude, while beam-related backgrounds, such as pion contamination, can be efficiently suppressed utilizing the High-energy Fragment Separator (HFRS)~\cite{10365405,Sheng:2023ojn} along with TOF detectors for Particle Identification (PID), thereby improving the beam purity. In this work, we consider a high-purity continuous muon beam.

\begin{figure}[htbp]
    \centering
    \includegraphics[width=0.7\linewidth]{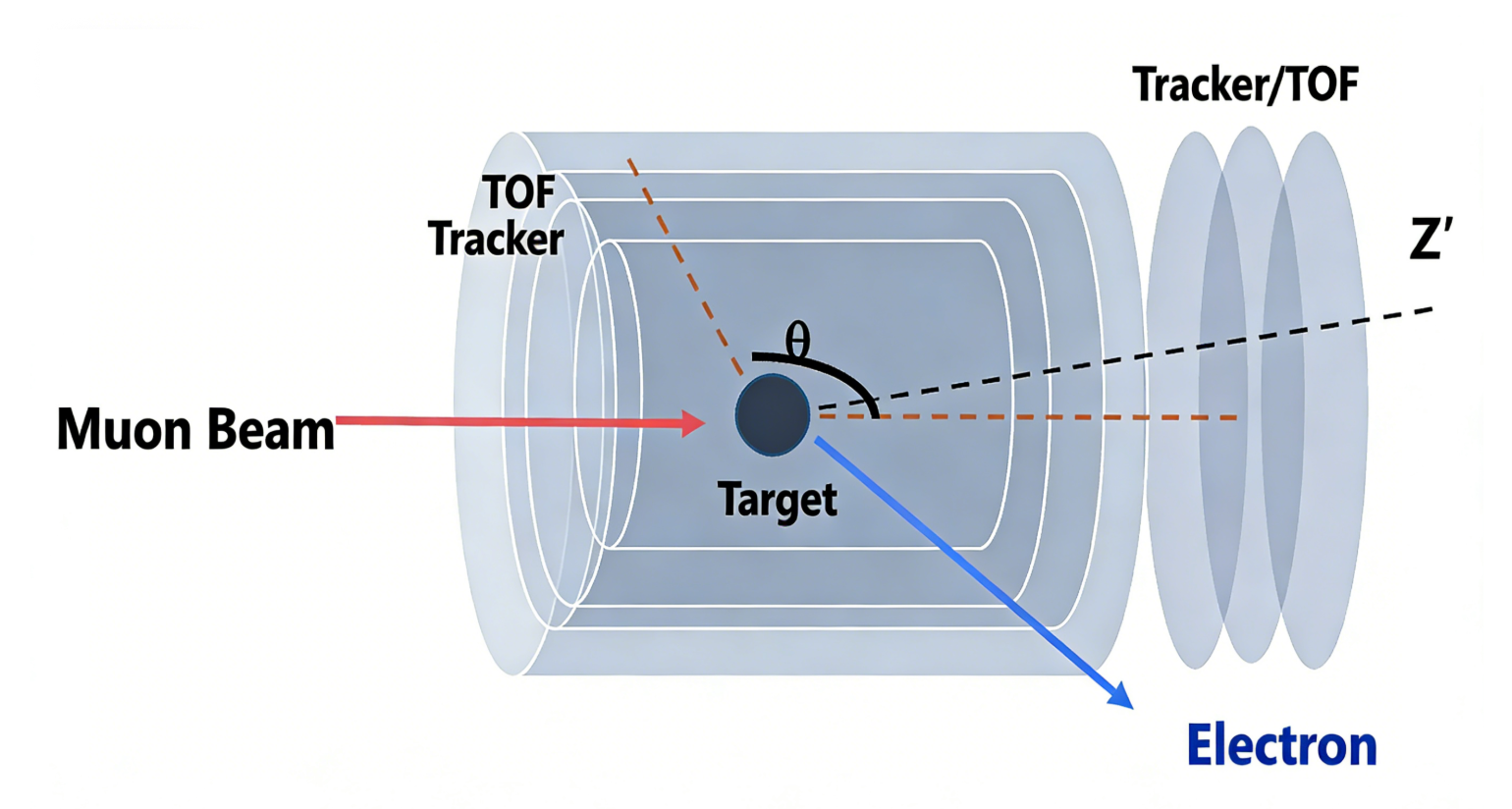}
    \caption{Schematic illustration of the DREAMuS detector.A proposed muon experiment to probe flavor violation mediator through muon on-target collisions}
    \label{fig:detector}
\end{figure}

In the proposed muon–nucleus scattering scenario, the signal is characterized by a flavor-violating interaction with a single recoil electron. It features a recoil electron with sizable transverse momentum and no additional charged particles at the interaction vertex.
To distinguish the kinematic features of muon-philic dark matter signals from those of SM particles, a high-precision detector is required. The schematic layout of the DREAMuS detector is shown in Fig.~\ref{fig:detector}. The detector consists primarily of a tracking system and a TOF system, designed to provide precise measurements of the position, momentum, and timing of charged particles produced at the target.

The tracking system adopts a cylindrical barrel geometry surrounding the target and consists of silicon strip detectors arranged in three concentric layers. Additional silicon strip layers are installed in both endcap regions to ensure coverage of forward and backward scattering particles. Each tracking layer is followed by an outer TOF layer, enabling precise timing measurements associated with reconstructed tracks.

The barrel has a total length of approximately 6~m and a radius of about 1.7\,m, providing angular acceptance for charged particles up to scattering angles of roughly 120 degree. A thin lead target is placed at the geometric center of the barrel. The TOF system achieves a time resolution of about 30\,ps, allowing accurate velocity measurements and contributing to background suppression through precise time correlation.

Three additional standalone tracking and TOF stations are installed along the beam direction, with a spacing of approximately 0.5\,m between successive layers. These stations serve as a dedicated veto system for the beam remnant.

\section{Event Generation and Simulation}

\subsection{Signal production}
\begin{figure}[h]
    \centering  \includegraphics[width=0.48\linewidth]{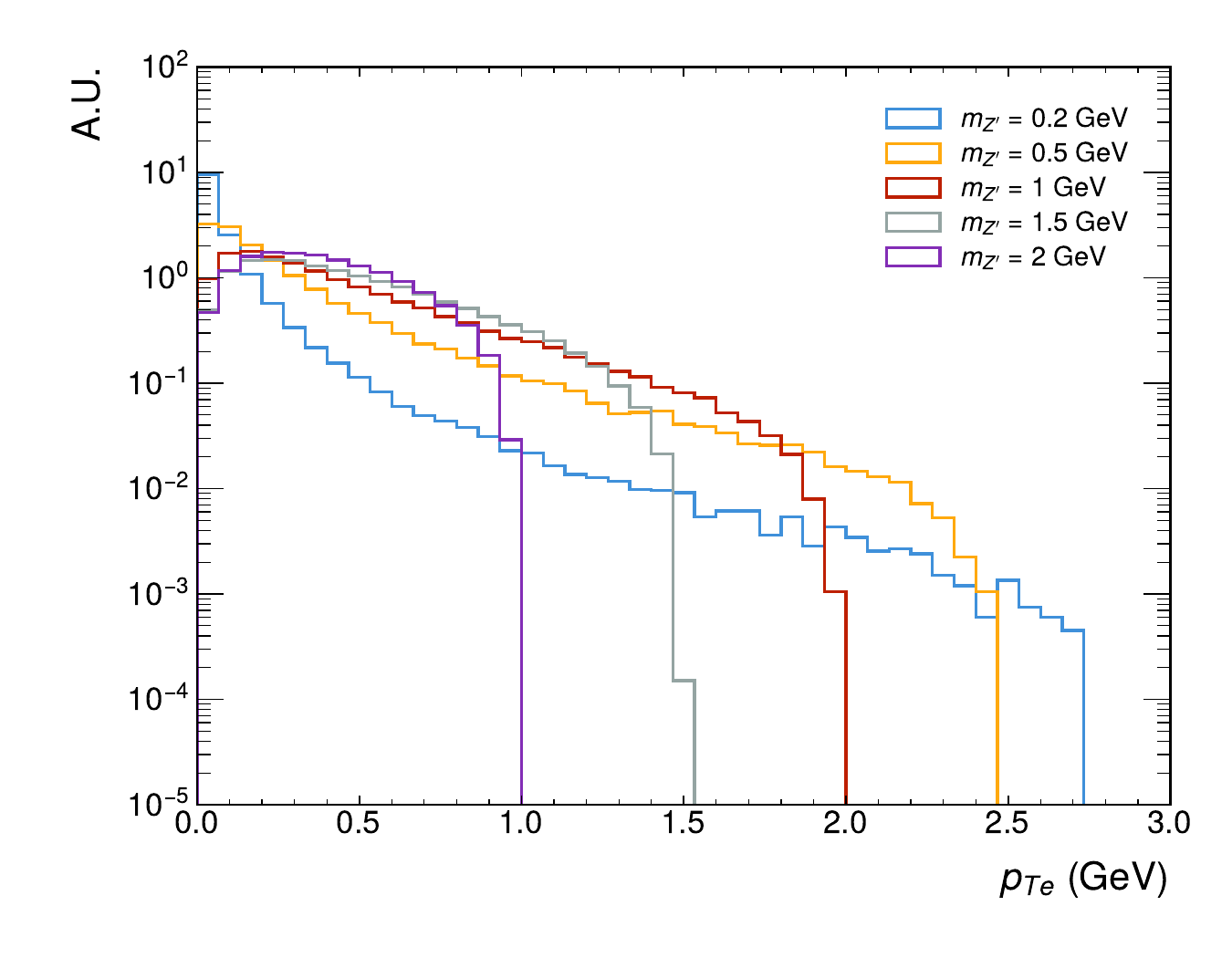}  \includegraphics[width=0.48\linewidth]{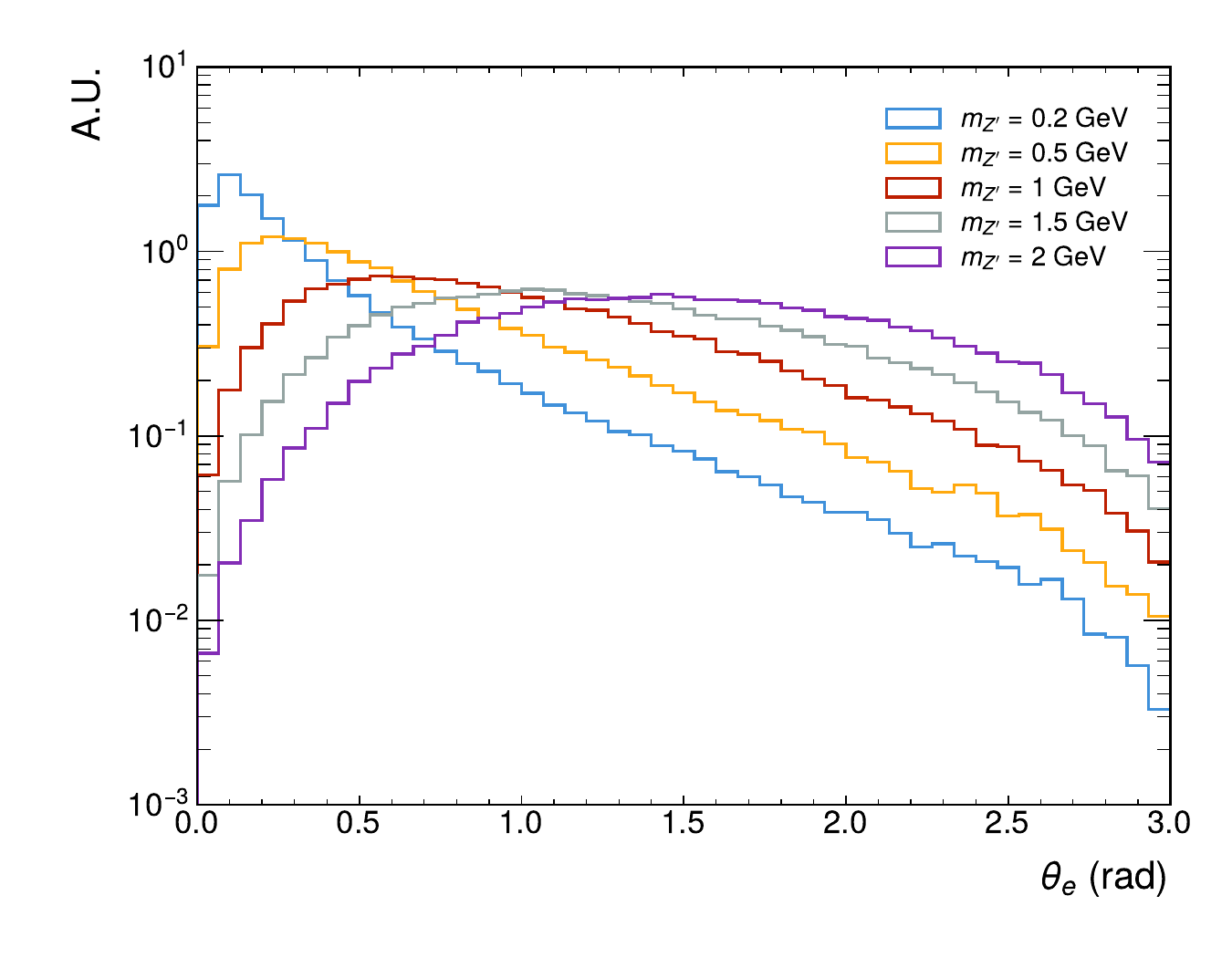}
    \caption{The transverse momentum ($p_{T}$) and scattering angle of the outgoing electron ($\theta_{e}$) for the $\mu N \rightarrow e N Z'$}
\label{fig:signalzprime_simu}
\end{figure}

\begin{figure}[h]
    \centering     \includegraphics[width=0.48\linewidth]{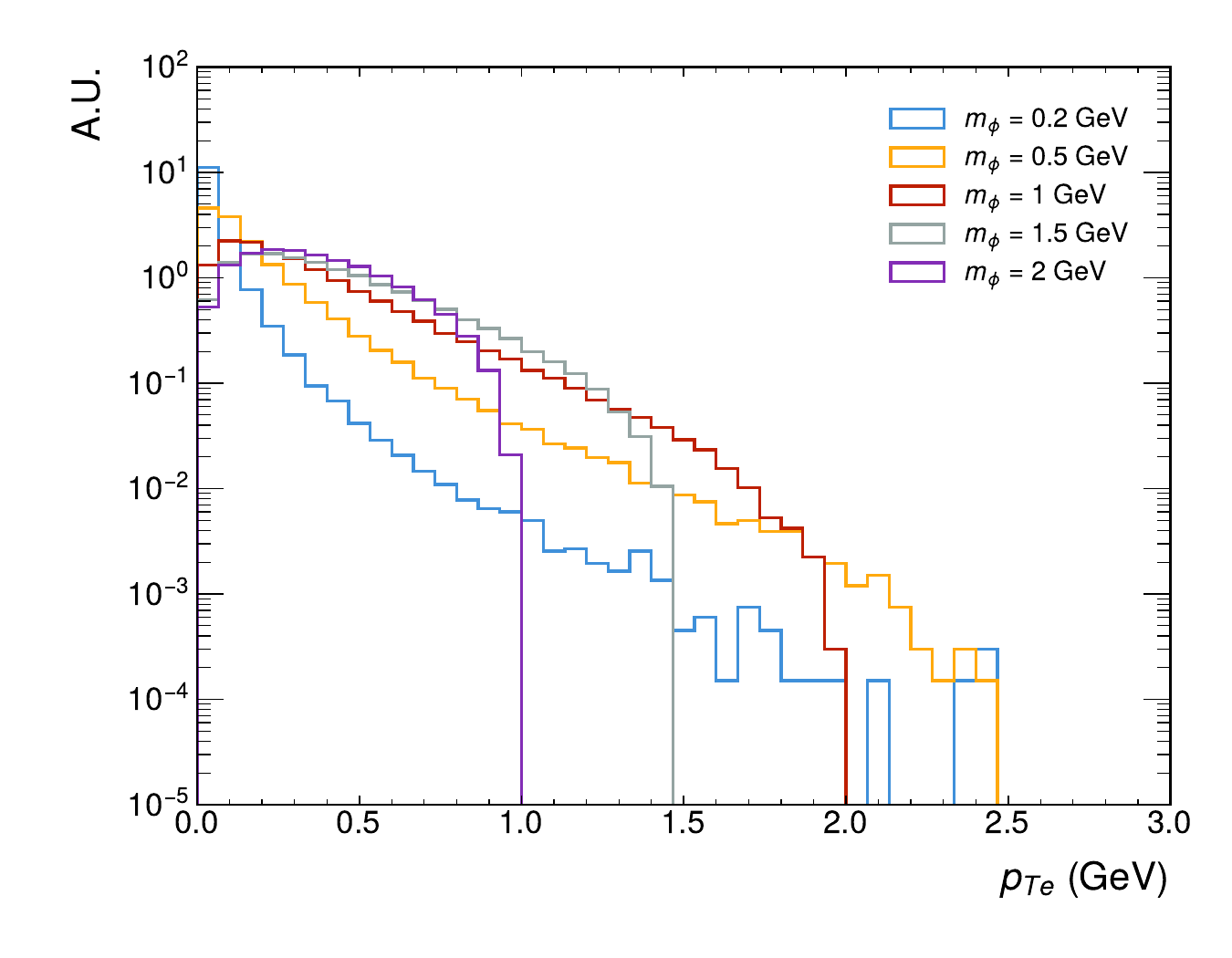} \includegraphics[width=0.48\linewidth]{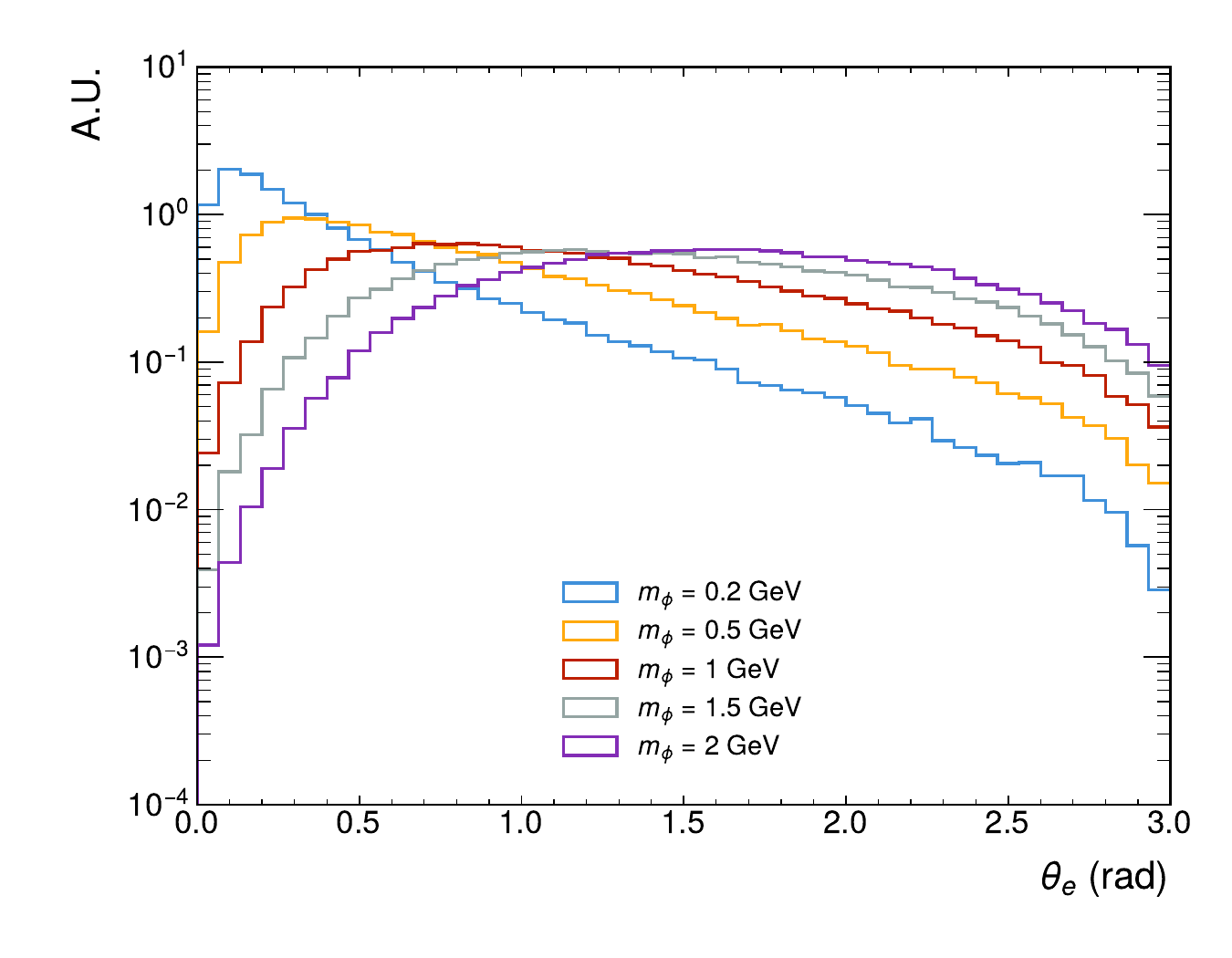}
    \caption{The transverse momentum ($p_{T}$) and scattering angle of the outgoing electron ($\theta_{e}$) for the $\mu N \rightarrow e N \phi$}
    \label{fig:signalphi_simu}
\end{figure}

CalcHEP v3.8.5~\cite{Belyaev_2013} is used to calculate the cross sections for both the radiation channel, $\mu N\rightarrow e N X$, and the annihilation channel, $\mu^{+}e^{-}\rightarrow X$, and to derive the kinematic distributions of the final states.
The signal samples are generated with an incident $\mu^{-}$ or $\mu^{+}$ beam energy of 3\,GeV, scanning the mediator mass $m_X$ from 106\,MeV to 2\,GeV. A total of 15 mass points are considered, with $10^5$ events generated for each point. The form factor of the target is set to be point-like. 
For lighter $m_{X}$, the recoil electron typically carries a larger transverse momentum $p_T$, while a heavier mediator results in a softer $p_T$ spectrum.
The angular distribution of the outgoing electron also shows a clear mass dependence, with the mean scattering angle $\theta_e$ increasing as $m_{Z'}$ becomes larger. A similar behavior is observed for the scalar mediator $\phi$ case. Representative distributions are shown in Fig.~\ref{fig:signalzprime_simu} for the $Z'$ scenario and in Fig.~\ref{fig:signalphi_simu} for the $\phi$ scenario.

The annihilation channel $\mu^{+} e^{-} \rightarrow X$ is also considered.Since this process produces only invisible final states, it requires a disappearance-style selection based on vetoing visible activity. 

\subsection{Background Estimation}
In this experiment, the dominant backgrounds originate from muon decays and muon interactions with target nuclei. Muon–nuclear processes are simulated using GEANT4~10.6.0~\cite{AGOSTINELLI2003250}, while muon decays are modeled with McMule~\cite{ulrich2025mcmulemontecarlo}. The backgrounds for the radiation channel and the annihilation channel are discussed separately.


\subsubsection{Backgrounds to the radiation channel}
\begin{itemize}
\item \textbf{Muon decay} 
Muon decays constitute an important background because their final states can mimic the signal. In particular, the three-body decay $\mu^- \to e^- \bar{\nu}_{e} \nu_{\mu}$ produces a single electron accompanied by missing energy. Rare decay modes, such as $\mu^- \to e^- \bar{\nu}_e \nu_\mu \gamma$ and $\mu^- \to e^- e^+ e^- \bar{\nu}_e \nu_\mu$ can also contribute when photons or charged leptons escape detection or are not fully reconstructed. Although these processes are suppressed by their small branching fractions, they may still yield non-negligible background contributions because of the high muon-beam intensity. Since muon decays in the rest frame are constrained by momentum conservation, the transverse momentum $p_{T}$ of the final-state electrons is typically small ($<50\,\mathrm{MeV}$). Fig.~\ref{fig:muon_decay} shows the energy and $p_T$ distributions of the final-state electrons for both the signal and background, where the signal electrons tend to have larger transverse momentum.

\begin{figure}[!htbp]  
    \centering  
    \includegraphics[width=0.48\linewidth]{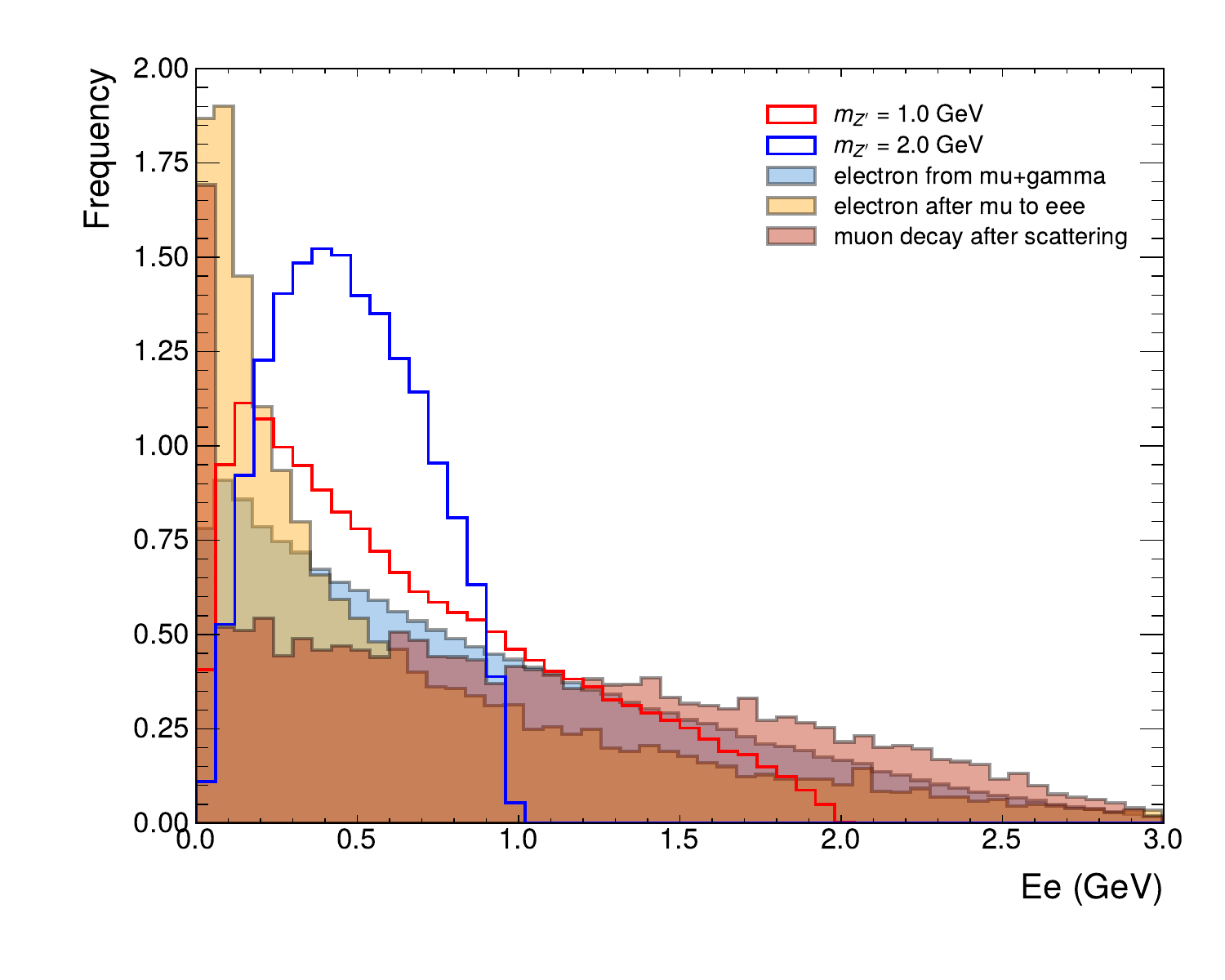}  
    \includegraphics[width=0.48\linewidth]{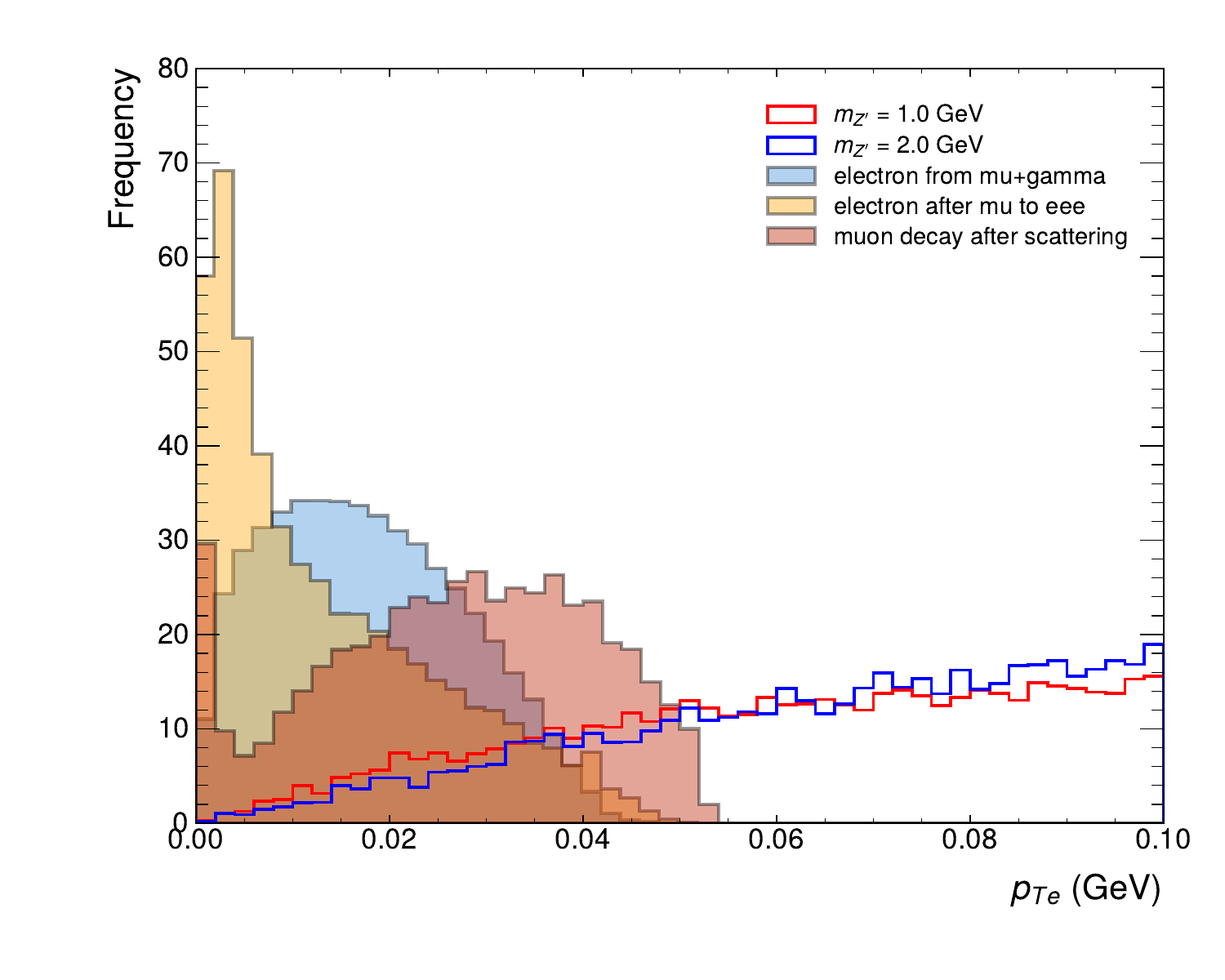}  
    \caption{Distributions of final-state electrons in energy and transverse momentum for muon decay and signal processes.The muon decay process is simulated by the McMule~\cite{ulrich2025mcmulemontecarlo}}  
    \label{fig:muon_decay}  
\end{figure}  


\item \textbf{Backgrounds from muon interactions with the target}
In addition to muon decays, background events can also arise from muon interactions with the target material. $\mu$-e scattering and ionization may produce recoil electrons that mimic the signal signature. Moreover, if a muon decays after Coulomb or elastic scattering, the decay electron may exhibit kinematic features similar to those of the signal. In high-$Z$ target materials, muon bremsstrahlung can also occur, with the emitted photons subsequently converting into $e^+e^-$ pairs. Muon--nucleus interactions may further contribute to the background, although such processes are often accompanied by additional final-state particles.

These background processes typically involve more than one charged particle with larger scattering angles. Based on the final-state particle identification, TOF information, and combined selections with the electron's transverse momentum $p_T$ and scattering angle $\theta$, these backgrounds can be effectively suppressed. 

\end{itemize}

\subsubsection{Backgrounds to the annihilation channel}

For the annihilation signal $\mu^+e^-\to X$, the dominant irreducible backgrounds are the SM processes $\mu^+e^-\to\nu_\mu\bar{\nu}_e$ and $\mu^+e^-\to\nu_\mu\bar{\nu}_e\gamma$, with cross sections of $5.5\times10^{-2}\,fb$ and $2.5\times10^{-3}\,fb$, respectively. These processes lead to fully invisible final states that are indistinguishable from the signal in the DREAMuS setup. Since no visible final-state particles are produced, they cannot be suppressed by detector-based selections and are therefore treated as irreducible backgrounds in the sensitivity estimation. For $10^{12}$ MOT, they correspond to about 0.1 background events in total.

\section{Event Selection}
\subsection{Radiation Channel}
\begin{figure}[htbp!]
    \centering
    \includegraphics[width=0.7\linewidth]{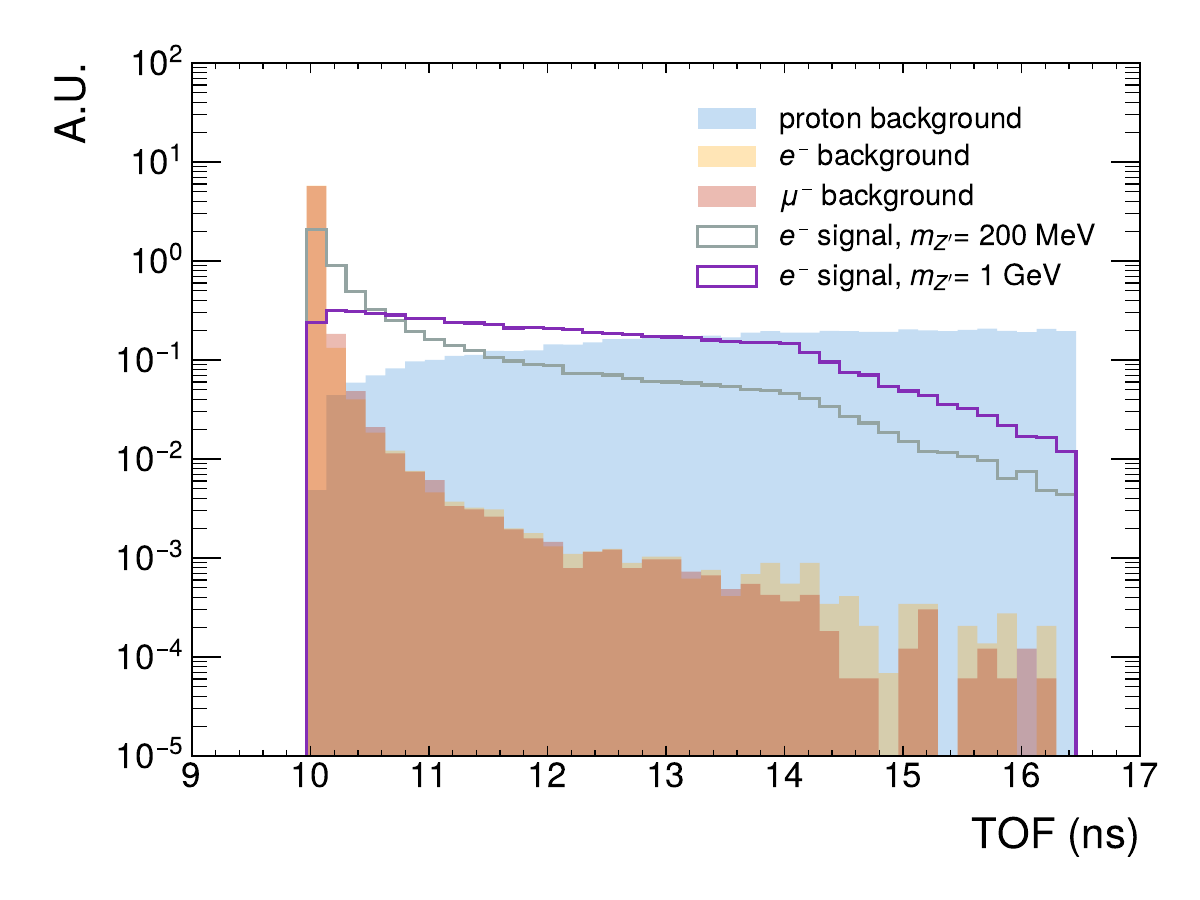}
    \caption{The time-of-flight(TOF) distribution for different particle}
    \label{fig:tof_distribution}
\end{figure}

In this channel, we simulate the main background processes, including muon-electron elastic scattering, muon ionization, bremsstrahlung, muon-nuclear interactions, and electron-pair production. To suppress these backgrounds, the following event-selection criteria are applied:
\begin{itemize}
    \item \textbf{Single track selection:} Events with only one reconstructed charged particle track are retained. This suppresses events are particularly from muon nuclear processes or electron pair production.
    \item \textbf{TOF selection:} The reconstructed track must have TOF in the window \( 11\,\mathrm{ns} < \mathrm{TOF} < 14\,\mathrm{ns} \). The distribution of TOF for different particle are shown in Fig.~\ref{fig:tof_distribution}, which shows that most hadrons have longer TOF while muons and electrons have smaller TOF, especially for muon beam remnant. 
    \item \textbf{Electron angular selection:} The electron candidate must have polar angle \( \theta > 43^\circ \), targeting signal electrons emitted at larger angles. 
    \item \textbf{Transverse momentum selection:} To suppress muon decay and elastic scattering backgrounds, the electron is required to have \( p_T > 20\,\mathrm{MeV} \).
\end{itemize}


Table~\ref{tab:bkg_cutflow} summarizes the number of background events surviving each selection. Muon decay is considered in each processed. After the full selection, no background events remain in the simulated samples, indicating a strong background rejection capability. The signal efficiency after all selections ranges from 10\% to 30\%, depending on the $Z'$ or $\phi$ mass.


\begin{table}[htbp]
\centering
\caption{Cutflow for the dominant background processes in the radiation channel.}
\label{tab:bkg_cutflow}
\resizebox{\linewidth}{!}{
\begin{tabular}{l|ccccc}
\toprule
\textbf{Process} & \textbf{Generated} & \textbf{Single track} & \textbf{TOF} & $\theta_e > 43^\circ$ & $p_T^e > 20$ MeV \\
\midrule
$\mu$--$e$ Scattering     & 50000  & 264   & 0 & 0 & 0 \\
$\mu$ Ionization          & 63089  & 2263  & 84& 0 & 0 \\
$\mu$ Bremsstrahlung      & 164411 & 135506&314&0 & 0 \\
e Pair-production        & 99281  & 138   & 0 & 0 & 0 \\
Elastic Scattering        & 113936 & 88917 & 37&20&0 \\
$\mu$--N Interaction      & 167120 & 144   & 3 & 2 & 0 \\
\bottomrule
\end{tabular}
}
\end{table}


The results demonstrate that the combination of single-track, TOF, and electron kinematic selections efficiently suppresses all SM backgrounds. In particular, the TOF and angular selections play a key role in rejecting electrons from elastic scattering and bremsstrahlung processes. No events survive after the full selection in the simulation, suggesting that the background contribution in the signal region is negligible with the current MC statistics.

\subsection{Annihilation Channel}
 For the annihilation signal $\mu^+ e^- \to X$, the final state is fully invisible. The event selection is therefore designed to veto all visible downstream activity arising from beam-induced muon interactions while retaining maximal signal efficiency. A series of vetoes is applied to suppress such backgrounds:
\begin{itemize}
    \item \textbf{Charged-particle veto:} Events with reconstructed downstream charged tracks are rejected. The assumed tracking efficiency is 99\%.
    \item \textbf{Tracker veto:} Events with any additional hits in the downstream tracker are rejected.
\end{itemize}
After these selections, the annihilation signal efficiency is expected to remain high, while visible backgrounds are strongly suppressed, shown in Table~\ref{tab:positivemu_cutflow}. The remaining irreducible backgrounds are SM processes such as $\mu^+ e^- \to \nu_\mu \bar{\nu}_e$ and $\mu^+ e^- \to \nu_\mu \bar{\nu}_e \gamma$. In addition, reducible backgrounds can arise from residual beam contamination, upstream interactions, hard bremsstrahlung, muon--nuclear interactions, trident or pair-production processes with undetected secondaries, and neutral particles escaping detection. Under the high-purity beam condition, these beam-related and detector-related backgrounds are expected to be subdominant and are not expected to affect the main conclusions of this study.

\begin{table}[htbp]
\centering
\caption{Cutflow for the dominant background processes in the
positive muon annihilation channel.}
\label{tab:positivemu_cutflow}
\resizebox{\linewidth}{!}{
\begin{tabular}{l|ccc}
\toprule
\textbf{Process} & \textbf{Generated events} & \textbf{No track} & \textbf{No hits on tracker} \\
\midrule
$\mu$ Decay            & 200000  & 1800 & 0 \\
$\mu$-e Scattering    & 63089   & 26   & 0 \\
$\mu$ Bremsstrahlung  & 328806  & 2806 & 0 \\
e Pair-production     & 9999973 & 609  & 0 \\
Elastic Scattering    & 13936   & 118  & 0 \\
$\mu$--N Interaction  & 167120  & 4    & 0 \\
\bottomrule
\end{tabular}
}
\end{table}

These results indicate that the combination of charged-particle and tracker vetoes provides effective rejection of visible backgrounds while preserving high efficiency for the invisible annihilation signal, with only a limited irreducible background remaining.

\section{Results}
The HIAF facility is expected to deliver a high-intensity muon beam with fluxes of approximately $3.3\times10^{5}~\mu^{-}/\mathrm{s}$ and $7.0\times10^{5}~\mu^{+}/\mathrm{s}$ at a beam energy of $3~\mathrm{GeV}$ in continuous-beam mode. For an effective running time of about 4000 hours for the $\mu^{-}$ beam and 2000 hours for the $\mu^{+}$ beam, the total number of muons on target (MOT) incident on a $40\,X_0$ ($\sim22~\mathrm{cm}$) lead target is estimated to be $5\times10^{12}$.

\begin{figure*}[htbp]
    \centering
    \includegraphics[width=0.48\linewidth]{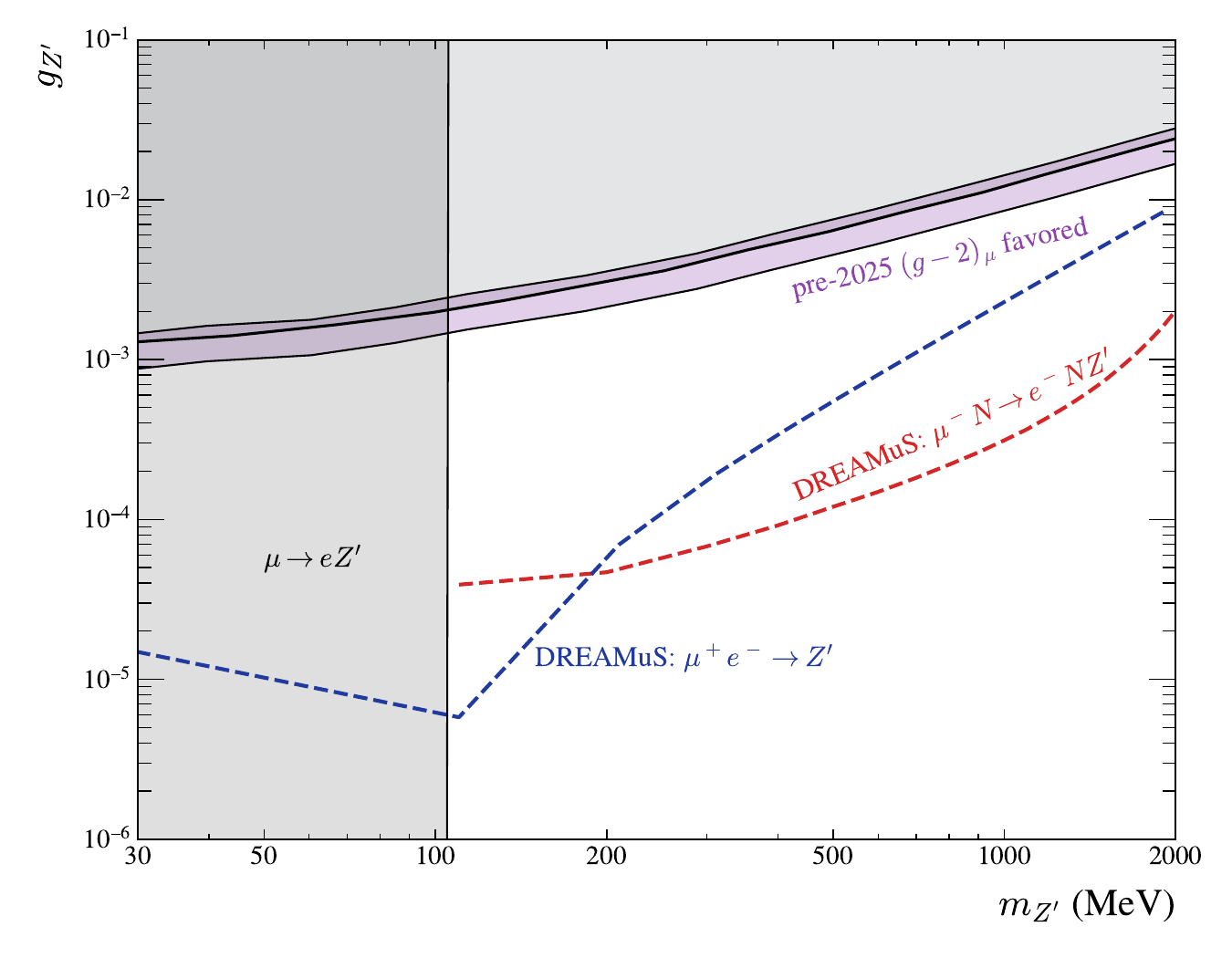}
    \includegraphics[width=0.48\linewidth]{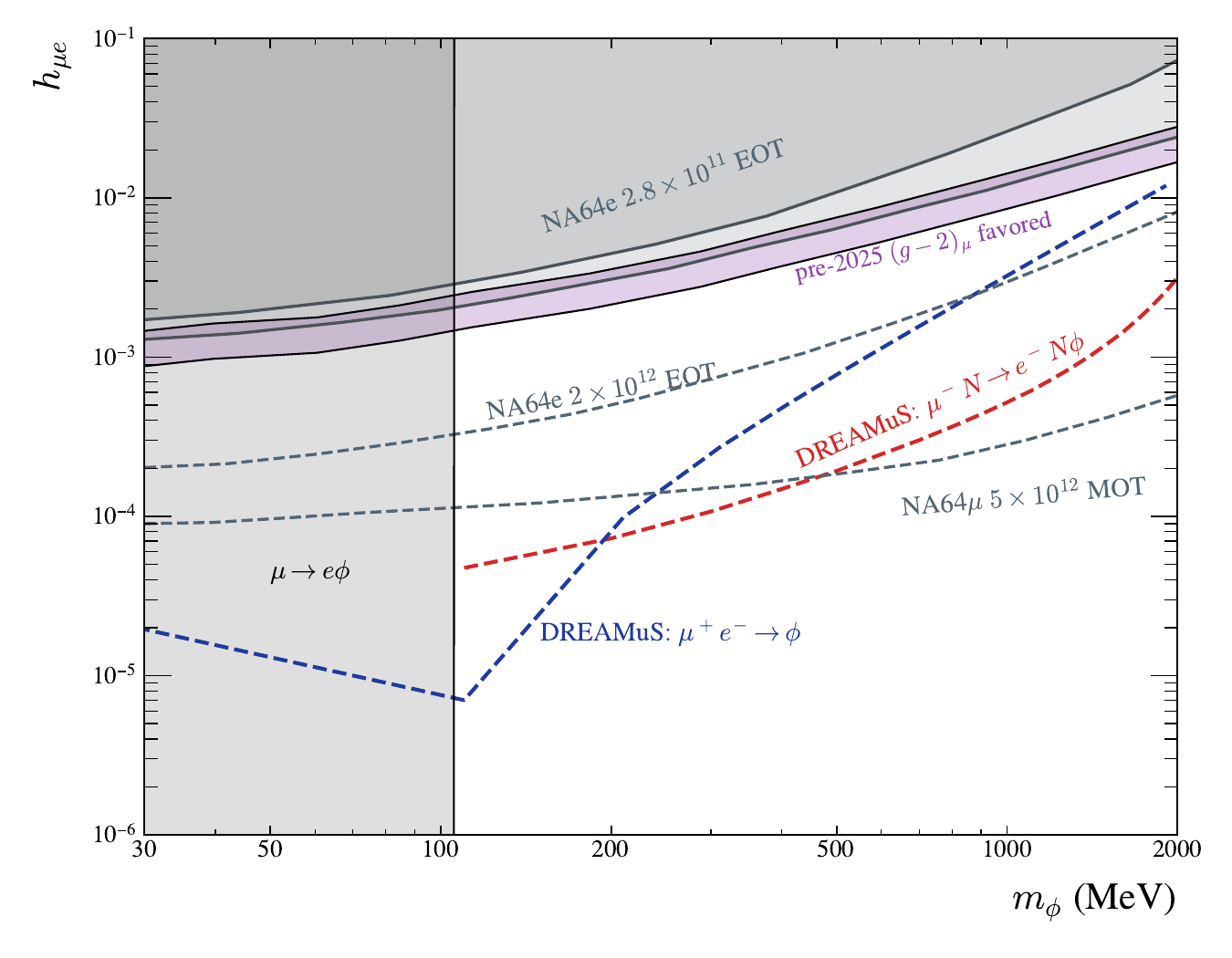}
    \caption{Projected DREAMuS sensitivity to flavor-violating mediator couplings at the 90\% C.L.. The left panel shows the expected reach for the vector mediator coupling $g_{Z'}$ as a function of the mediator mass $m_{Z'}$, while the right panel shows the corresponding reach for the scalar mediator coupling $h_{\mu e}$ in the $\phi$ scenario. In both panels, the red dashed curves correspond to the radiation channel and the blue dashed curves to the annihilation channel. The shaded regions are excluded by existing constraints~\cite{TWIST:2015mueX}, and the purple band indicates the parameter space favored by the pre-2025 $(g-2)_\mu$ anomaly~\cite{Aoyama:2020ynm,Muong-2:2021vma,Muong-2:2023cdq}. The projected sensitivities of NA64$_e$ and NA64$_\mu$ are taken from Refs.~\cite{NA64:2022rme,NA64:2024klw}.}
    \label{fig:dreamus_limit}
\end{figure*}

\begin{figure*}[htbp]
    \centering
    \includegraphics[width=0.48\linewidth]{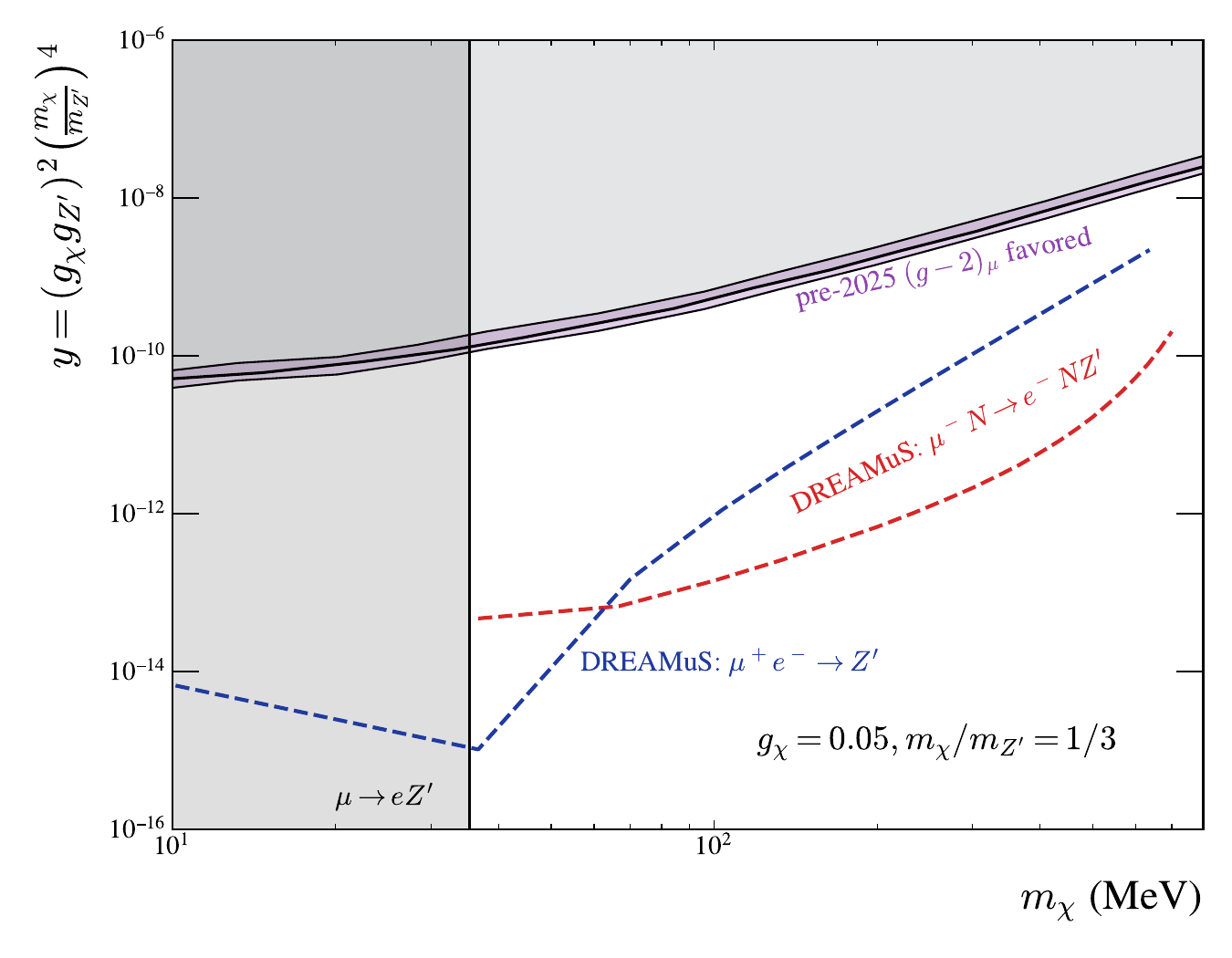}
    \includegraphics[width=0.48\linewidth]{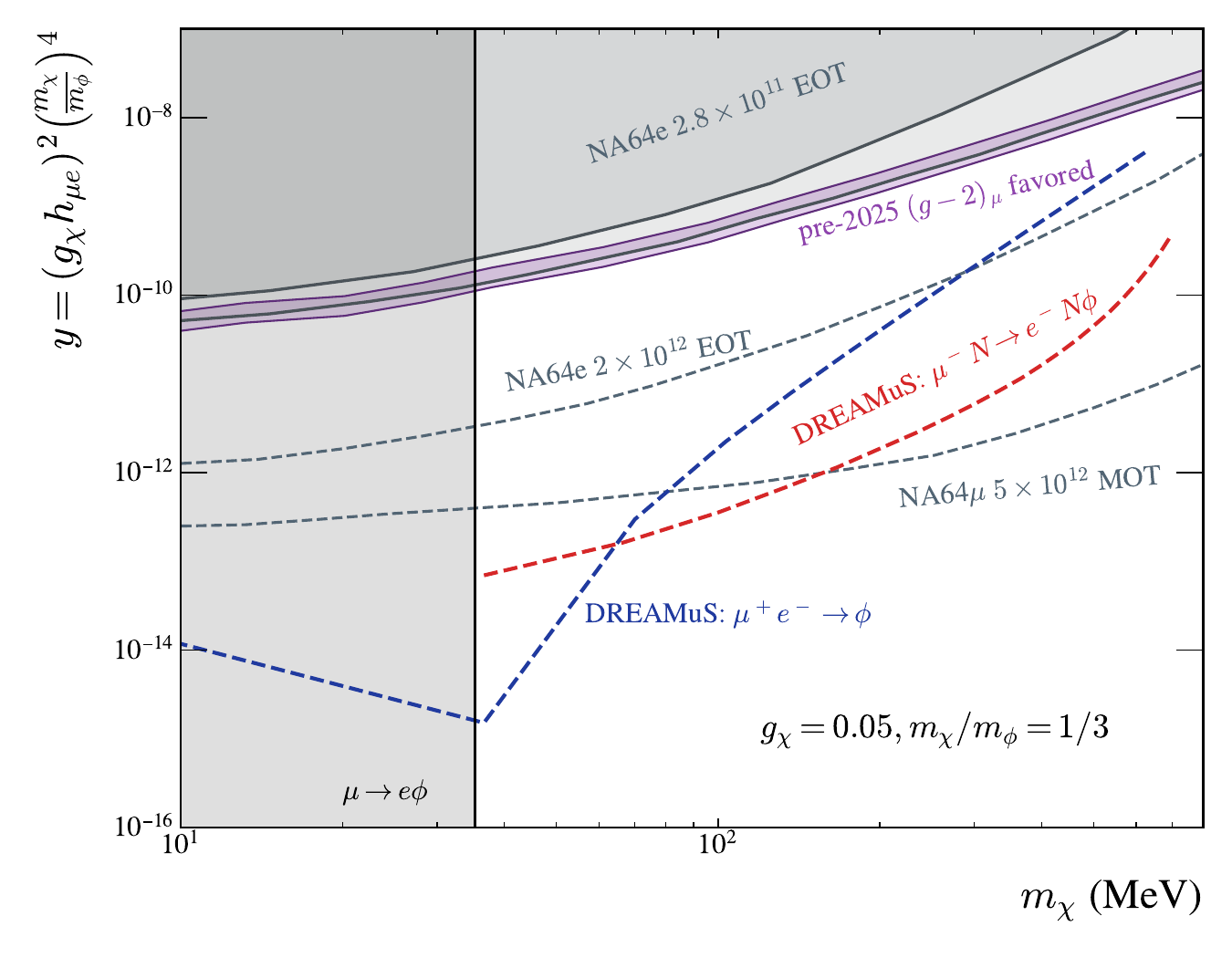}
    \caption{Projected DREAMuS sensitivity to flavor-violating mediator couplings at the 90\% C.L., expressed in terms of $y=(g_{\chi}g_{Z'})^{2}\left(\frac{m_{\chi}}{m_{Z'}}\right)^{4}$ and $y=(g_{\chi}h_{\mu e})^{2}\left(\frac{m_{\chi}}{m_{\phi}}\right)^{4}$. In both panels, the red dashed curves correspond to the radiation channel and the blue dashed curves to the annihilation channel. The shaded regions are excluded by existing constraints~\cite{TWIST:2015mueX}, and the purple band indicates the parameter space favored by the pre-2025 $(g-2)_\mu$ anomaly~\cite{Aoyama:2020ynm,Muong-2:2021vma,Muong-2:2023cdq}. The projected sensitivities of NA64$_e$ and NA64$_\mu$ are taken from Refs.~\cite{NA64:2022rme,NA64:2024klw}.}
    \label{fig:dreamus_limit_y}
\end{figure*}

After applying the event selection criteria described in the previous section, all considered SM backgrounds are fully suppressed within the available MC statistics for both the radiation and annihilation channels. The expected sensitivity is therefore evaluated under the assumption of a background-free search. Upper limits on the signal yield are set at the 90\% confidence level (C.L.) ~\cite{Cowan:2010js}. The corresponding limits on the coupling strengths $g_{Z'}$ and $h_{\mu e}$ are obtained from the signal yield, taking into account the experimental exposure and the total signal efficiency $\epsilon_{\mathrm{sig}}$.
Fig.~\ref{fig:dreamus_limit} shows the projected DREAMuS exclusion limits on the flavor-violating mediator couplings. The same results are recast in Fig.~\ref{fig:dreamus_limit_y} for the dark matter interpretation in terms of the effective coupling variables
$y = (g_\chi g_{Z'})^{2}\left(\frac{m_\chi}{m_{Z'}}\right)^{4}
$ and $y = (g_\chi h_{\mu e})^{2}\left(\frac{m_\chi}{m_{\phi}}\right)^{4},$
with $g_\chi=0.05$ and $m_\chi/m_X = 1/3$.
The left panel presents the expected exclusion limit on the vector mediator coupling $g_{Z'}$ as a function of the $m_{Z'}$, while the right panel shows the corresponding sensitivity to the scalar mediator coupling $h_{\mu e}$ in the $\phi$ scenario. The red dashed curves correspond to the radiation channel, whereas the green dashed curves represent the annihilation channel.
The radiation channel provides robust sensitivity over the medium to high mass region, from 200\,MeV to 1\,GeV. By contrast, the invisible annihilation channel dominates in the low-mass region, improving the sensitivity by more than an order of magnitude compared with the radiation channel alone.

Also shown in Fig.~\ref{fig:dreamus_limit} and Fig.~\ref{fig:dreamus_limit_y} are existing constraints and projected sensitivities from other experiments. The shaded regions indicate parameter space excluded by current bounds~\cite{ParticleDataGroup:2024cfk,HEINTZE1979365}, while the projected sensitivity curves from NA64$_e$~\cite{NA64:2022rme} and NA64$_\mu$~\cite{NA64:2024klw} are shown for comparison. The purple band denotes the parameter space favored by the pre-2025 $(g-2)_\mu$ anomaly at the 1$\sigma$ and 2$\sigma$ confidence levels.~\cite{Muong-2:2025xyk, Aliberti:2025beg, Beltran:2026ofp}. DREAMuS is expected to probe a significant fraction of this motivated parameter space, particularly in the low-mass regime where the invisible annihilation channel substantially extends the sensitivity beyond existing and projected limits.

Complementary information on light LFV mediators also comes from stopped-muon two-body decay searches. Existing limits from the TWIST experiment on $\mu^+\to e^+X^0$, based on $5.8\times10^8$ positive-muon decays, have been recast into direct limits on the mediator couplings and overlaid in Figs.~\ref{fig:dreamus_limit} and~\ref{fig:dreamus_limit_y}. These searches probe the decay-at-rest topology for $m_X<m_\mu-m_e$, covering much of the $13-80\,\mathrm{MeV}$ mass range~\cite{TWIST:2015mueX}. A complementary probe of light LFV mediators has been proposed using dedicated $\mu^+$ validation or decay-at-rest datasets at stopped-muon facilities such as Mu2e~\cite{Bartoszek:2015tdr} and COMET~\cite{Abramishvili:2020comet}, through searches for $\mu^+\to e^+X$ rather than the standard $\mu^-N\to e^-N$ conversion observable~\cite{Hill:2024mueX}. We do not overlay the projected Mu2e/COMET sensitivities, because they are prospective stopped-muon sensitivities restricted to $m_X<m_\mu-m_e$ and depend on facility-specific assumptions for the stopped-$\mu^+$ sample, acceptance, and background treatment. Complementary constraints on light LFV bosons have also been derived from electron beam-dump experiments, including E137-based studies of scalar and vector mediators with $\mu-e$ flavor-violating couplings~\cite{Araki:2021wum}. These probes are relevant by crossing symmetry, but the published limits are primarily obtained from searches for boson decays into $e^+e^-$ pairs and are therefore not directly overlaid on the present plots, which focus on invisible-decay dark-sector benchmarks.

\section{Summary and Outlook}
In this work, we explore a novel experiment, DREAMuS, to search for lepton-flavor-violating processes in the radiation channel, \(\mu N\rightarrow e N X\), followed by \(X\rightarrow\chi\chi\), and in the invisible channel, \(\mu^{+}e^{-}\rightarrow\chi\chi\), where a flavor-violating boson \(X\) is produced in association with an invisible state \(\chi\). The signal processes are generated with CalcHEP, and the backgrounds are simulated with GEANT4. The proposed setup is based on a high-intensity muon beam available at HIAF, operating at a beam energy of 3~GeV. A dedicated detector concept with a cylindrical geometry, precise tracking, and time-of-flight capabilities is considered. The characteristic kinematic and timing information of the recoil electron provides strong discrimination against background. As a result, the proposed experiment achieves competitive sensitivity to the coupling parameters \(g_{Z'}\) and \(h_{\mu e}\). For \(5\times10^{12}\) MOT on a 22~cm lead target, a 90\% C.L. limit is set on the coupling parameter, reaching the \(10^{-4}\) level in the radiation channel. Compared with the radiation channel, which has better sensitivity in the 200~MeV to 1~GeV region, the invisible channel achieves better sensitivity in the low-mass region around 100~MeV, with an improvement of about one order of magnitude, reaching close to \(10^{-5}\).

Future studies focus on full detector simulation, further optimization of the detector setup, more detailed background modeling, and the incorporation of possible systematic uncertainties. 
This work establishes a solid foundation for the DREAMuS experiment at HIAF and provides a new pathway to probe physics beyond the Standard Model.

\begin{acknowledgments}
We thank Tao Han for helpful comments and discussions.
This work was supported by National Key R\&D Program of China
(Grant Nos. 2024YFA1610600 and 2024YFA1610603); National Natural Science Foundation
of China (Grants No. 12475108, 12375101, 12425506); Research Program of State Key Laboratory of Heavy Ion Science and Technology, Institute of Modern Physics, Chinese Academy of Sciences(Grant No. HIST2025CS06).
\end{acknowledgments}
\nocite{*}

\bibliography{apssamp}

\end{document}